\documentclass[%
 reprint,
superscriptaddress,
 amsmath,amssymb,
 aps,
 prx,
]{revtex4-2}

\usepackage{xcolor}
\usepackage{url}

\usepackage{graphicx}% Include figure files
\usepackage{dcolumn}% Align table columns on decimal point
\usepackage{bm}% bold math
\begin{document}

\preprint{APS/123-QED}

\title{Floquet Mode Resonance: Trapping light in the bulk mode of a Floquet topological insulator by quantum self-interference}

\author{Shirin Afzal}
\email{safzal1@ualberta.ca}
\author{Vien Van}%
\affiliation{Department of Electrical and Computer Engineering, University of Alberta, Edmonton, AB, T6G 2V4, Canada}


\begin{abstract}
Floquet topological photonic insulators characterized by periodically-varying Hamiltonians are known to exhibit much richer topological behaviors than static systems.   In a Floquet insulator, the phase evolution of the Floquet-Bloch modes plays a crucial role in determining its topological behaviors.  Here we show that by perturbing the driving sequence, it is possible to manipulate the cyclic phase change of the system over each evolution period to induce quantum self-interference of a bulk mode, leading to a new topological resonance phenomenon called Floquet Mode Resonance (FMR).  The FMR is fundamentally different from other types of optical resonances in that it is cavity-less since it does not require physical boundaries. Its spatial localization pattern is instead dictated by the driving sequence and can thus be used to probe the topological characteristics of the system.  We demonstrated excitation of FMRs by edge modes in a Floquet octagon lattice on silicon-on-insulator, achieving extrinsic quality factors greater than $10^4$.  Imaging of the scattered light pattern directly revealed the hopping sequence of the Floquet system and confirmed the spatial localization of FMR in a bulk-mode loop. The new topological resonance effect could enable new applications in lasers, optical filters and switches, nonlinear cavity optics and quantum optics.  
\end{abstract}

\maketitle
%%%%%%%%%%%%%%%%%%%%%%%%%%  body  %%%%%%%%%%%%%%%%%%%%%%%%%%
\section{Introduction}
Topological photonic insulators (TPIs) provide a rich playground for exploring both the physics of periodic systems as well as applications of their exotic properties \cite{lu2014topological, ozawa2019topological}. In particular, Floquet TPIs characterized by periodically-varying Hamiltonians have recently gained much attention as they can exhibit richer topological behaviors than static undriven systems, such as the existence of anomalous Floquet insulator (AFI) edge modes in lattices with topologically trivial energy bands \cite{rechtsman2013photonic, yang2020photonic, maczewsky2017observation, mukherjee2017experimental, peng2016experimental, gao2016probing, afzal2020realization}.  In a periodically-driven system, the evolution of the phase bands of the Floquet-Bloch modes over each period plays a crucial role in determining its topological behaviors \cite{Nathan2015,Nakagawa}. Here we investigate the possibility of manipulating the cyclic phase change of a Floquet mode to induce quantum interference in the lattice bulk, leading to a new resonance phenomenon as well as a new mechanism for probing the topological behaviors of Floquet TPIs.

The ability to form robust high quality factor resonators in a topological lattice is of practical interest as it would significantly broaden the range of applications of TPIs such as in lasers, filters, nonlinear cavity optics and quantum optics 
\cite{ota2020active,  siroki2017topological, he2019topological,  smirnova2020nonlinear, barik2020chiral, xie2020cavity}.  In 2D TPI lattices, travelling-wave resonators can be realized by exploiting the confinement of edge modes at the interface between topologically trivial and nontrivial insulators to form ring cavities, although these tend to have very long cavity lengths as they require many lattice periods  \cite{bahari2017nonreciprocal,bandres2018topological,  zeng2020electrically, yu2020critical, yang2018topological}.
Topological resonators can also be realized by creating line defects \cite{li2018topological} or point defects in the lattice bulk, e.g., by spatially shifting air holes in a photonic crystal to create a Dirac-vortex topological cavity \cite{gao2020dirac}.  The resonance mode is pinned to the midgap and can be regarded as the 2D counterpart of 1D resonance modes in  distributed feedback lasers \cite{haus1976antisymmetric} and vertical cavity surface emitting lasers (VCSELs) \cite{chuang2012physics}. In another variant of defect mode cavities \cite{shao2020high}, resonant confinement occurs due to band inversion-induced reflections from the interface walls as a result of the different parity modes inside and outside the cavity.  The cavity mode is a bulk mode located at the $\Gamma$ point of the energy band diagram very close to the edge of the topological bandgap. Recently, it was shown that topological corner states with zero energy can also be used to form resonances in a TPI \cite{noh2018topological, ota2019photonic, mittal2019photonic2, zhang2020low}. However, the mode is not tunable and can only be formed at the corners of the lattice.  Indeed all the topological photonic resonators reported to date are not continuously tunable and have only been realized for static TPI systems.

\begin{figure*}[t]
     \centering
     \includegraphics[width=1\linewidth]{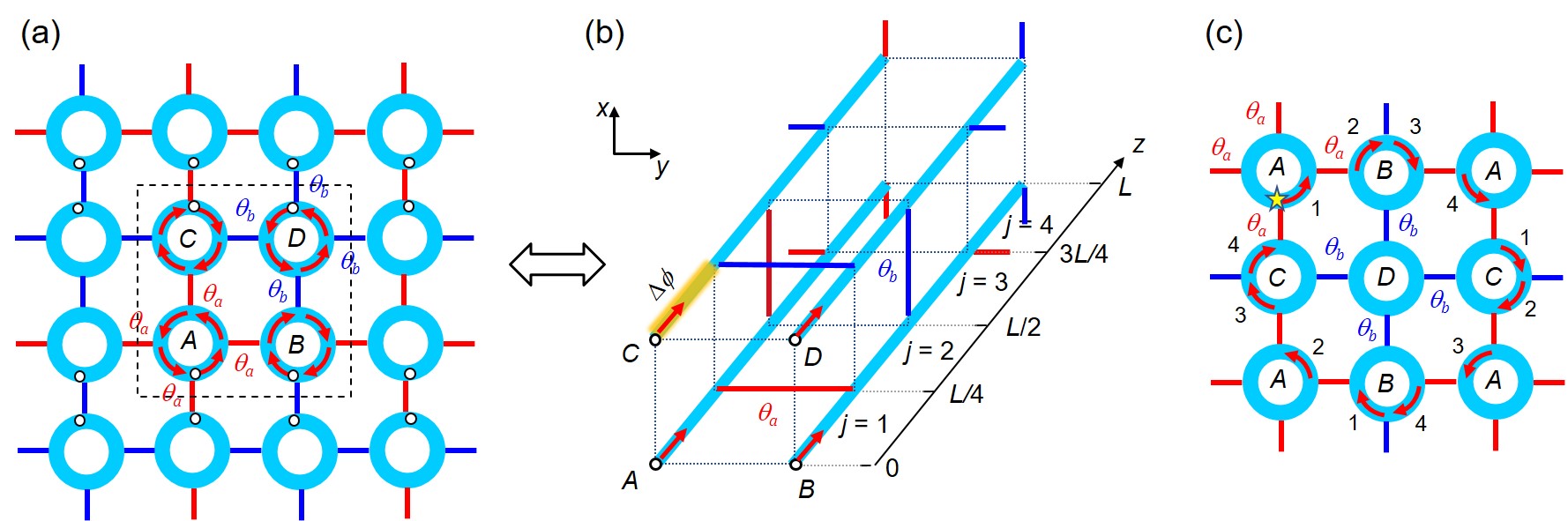}
     \caption{Driving sequence of a 2D Floquet microring lattice.  (a) Schematic of the lattice showing a unit cell with four microrings $\{A, B, C, D\}$ and coupling angles $\theta_a > \theta_b$. (b) Equivalent coupled-waveguide array representation of the microring lattice, obtained by cutting the microrings at the points indicated by the open circles in (a) and unrolling them into straight waveguides.  The system evolves periodically in the direction of light propagation $z$ in each microring, with each period consisting of four coupling steps $j = \{1, 2, 3, 4\}$. Also shown is a phase detune $\Delta\phi$ applied to microring $C$ in step $j = 1$ to perturb the drive sequence. (c) Spatial localization of a bulk mode in a loop (red arrows): starting from step $j = 1$ in microring $A$ (yellow star), the hopping sequence of the lattice guarantees that light returns to its initial point after 3 evolution periods.}
     \label{fig:1}
 \end{figure*}
 
Here we report a new resonance mechanism in a Floquet TPI whereby light is trapped in a Floquet bulk mode due to quantum self-interference. The resonance effect, which we refer to as Floquet Mode Resonance (FMR), is achieved by adiabatically tuning the drive-induced phase of a Floquet mode to achieve constructive interference. This has the concomitant effect of shifting its quasienergy into a topological bandgap to form an isolated flat-band state that is spatially localized in a bulk-mode resonant loop. The FMR is fundamentally different from other types of topological resonances in that it is cavity-less since it does not require creating physical boundaries in the lattice. The lack of scattering from interface discontinuities means that FMRs can potentially have very high Q factors.  The resonance can be formed anywhere in the lattice bulk and can be continuously tuned across a topological bandgap.  In addition, the spatial localization pattern of the FMR is dictated by the driving sequence of the Floquet TPI and can thus be used to probe the topological behaviors of the system.  We demonstrated FMR in a Floquet octagon lattice realized in silicon-on-insulator (SOI), using thermo-optic heater to excite and tune the resonance. Imaging of the scattered light intensity pattern reveals the hopping sequence of the Floquet mode and provides direct evidence of FMR localized in a bulk-mode loop. Our work not only introduces a versatile way for forming high-Q resonances in a Floquet lattice, but also provides a new method for probing the topological behaviors of bulk modes in Floquet systems.

\section{Theoretical Origin of FMR}
 
\begin{figure*}[htp]
     \centering
     \includegraphics[width=1\linewidth]{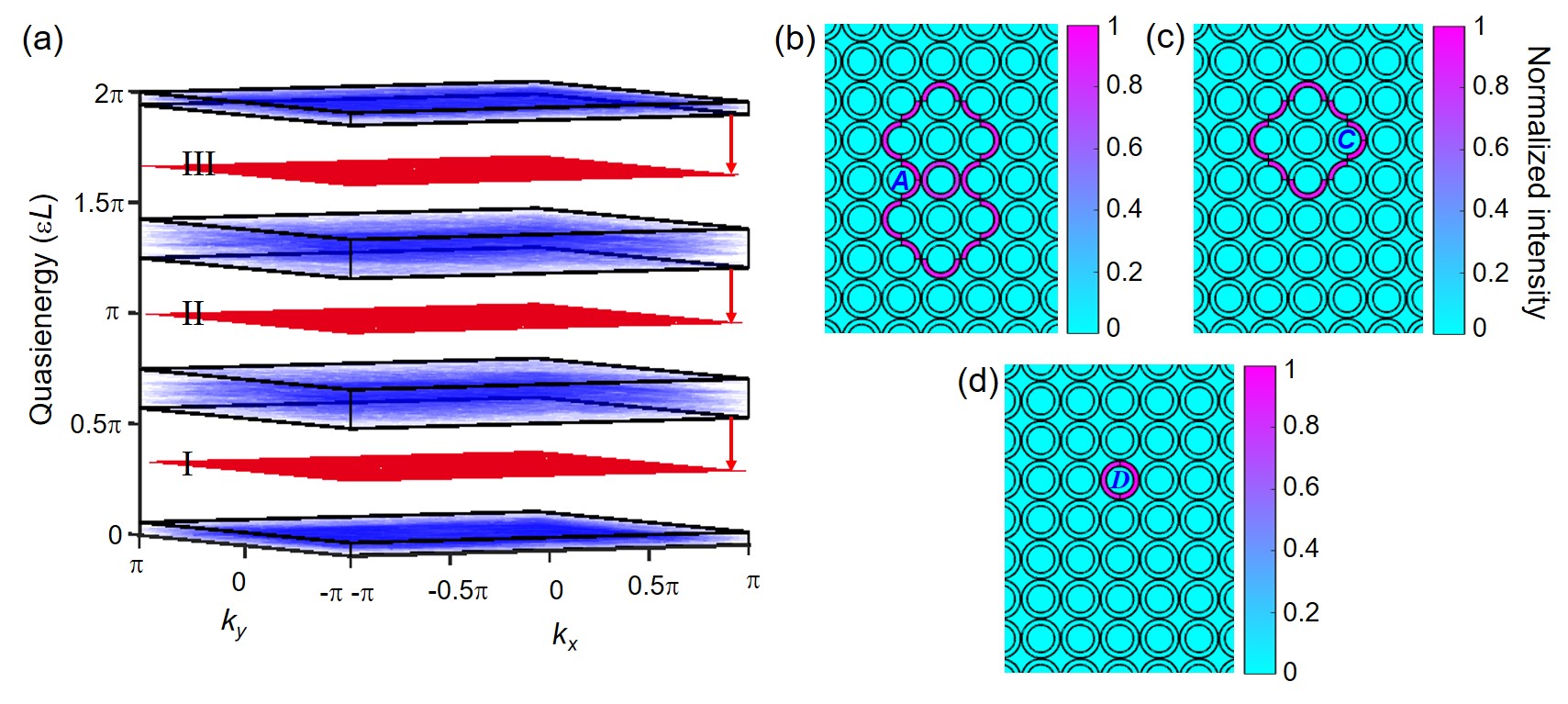}
     \caption{Frequency and spatial localizations of energy-shifted Floquet states. (a) One Floquet Brillouin zone of the quasienergy band diagram of an AFI microring lattice consisting of $5\times5$ unit cells with $\theta_a = 0.458\pi$ and $\theta_b=0.025\pi$. The blue bands are composite transmission bands of Floquet states separated by three topological bandgaps (labeled I, II and III). The red bands are the flat bands of Floquet bulk modes which are lifted from the transmission band manifolds due to a phase detune $\Delta\phi = \pi$ applied to step $j=1$ of microring $C$. (b)-(d)
      Intensity distributions of the energy-shifted bulk modes $\Phi_s$ obtained when phase detune $\Delta\phi = \pi$ is applied to segment $j=1$ of microring $A$, $C$ and $D$, respectively.} 
     \label{fig:2}
 \end{figure*}

 The Floquet TPI we consider is a 2D square lattice of coupled microring resonators with identical resonance frequencies. The lattice is characterized by two different coupling angles ($\theta_a > \theta_b$) in each unit cell (Fig.~\ref{fig:1}(a)), where $\theta_a$ ($\theta_b$) represents the strong (weak) coupling between resonator $A$ ($D$) and its neighbors.  Although 2D microring lattices have been shown to exhibit Chern insulator behavior associated with static systems \cite{hafezi2011robust, hafezi2013imaging, mittal2019photonic,leykam2020topological}, the existence of FMRs can only be predicted by treating the system as a Floquet insulator with periodically-varying Hamiltonian.  As light circulates around each microring, it couples periodically to its neighbors, so that the Bloch modes of the lattice evolve in a cyclical motion with a period equal to the microring circumference $L$. The quasienergy spectrum of the lattice thus has a periodicity of $2\pi/L$. Within each Floquet Brillouin zone, the microring lattice in general has three bandgaps (Fig.~\ref{fig:2}(a)), which can exhibit different topological phases, including AFI behavior, depending on the coupling angles ($\theta_a, \theta_b$) \cite{afzal2018topological}. To better elucidate the Floquet nature of the lattice, we transform it into an equivalent 2D array of periodically-coupled waveguides (Fig.~\ref{fig:1}(b)) \cite{afzal2018topological}, with each period consisting of 4 coupling steps between different pairs of adjacent waveguides.  In the limit of perfect coupling $(\theta_a = \pi/2, \theta_b = 0)$, the hopping sequence guarantees that light starting from site $A$ in a unit cell will return to its position after 3 periods, tracing out a bulk-mode loop depicted in Fig.~\ref{fig:1}(c). However, in a uniform Floquet lattice, such a bulk mode does not exist in a bandgap since its phase change around the loop is not equal to an integer multiple of 2$\pi$.

 Suppose that we now perturb the driving sequence by introducing a phase shift $\Delta\phi$ in coupling step $j$ of a microring in the lattice (Fig.~\ref{fig:1}(b)).  Taking a block of $N \times N$ unit cells with the perturbed microring located near its center, for sufficiently large $N$, we can treat this block as a supercell of an infinite periodic lattice. Using the coupled-waveguide array model, we can write the equation of motion of the supercell as
 \begin{eqnarray}
     i\frac{\partial}{\partial z} |\psi(\textbf{k},z)\rangle  &=& H(\textbf{k},z)|\psi(\textbf{k},z)\rangle\nonumber \\
     &=& \sum\limits_{j=1}^4[H_{FB}^{(j)}(\textbf{k},z) + H_D^{(j)}]|\psi(\textbf{k},z)\rangle
\end{eqnarray}
where $\textbf{k}$ is the crystal momentum in the $x$-$y$ plane, $H_{FB}^{(j)}$ is the Floquet-Bloch Hamiltonian of the unperturbed supercell in step $j$ \cite{afzal2018topological}, and $H_D^{(j)}$ is the perturbed Hamiltonian in the same step.  The perturbed Hamiltonian matrix is zero everywhere except for a term of $-\Delta\phi/L$ in its $k^\textrm{th}$ diagonal element corresponding to the detuned microring $k$.  Any state of the system evolves as $|\psi(\textbf{k},z)\rangle = U(\textbf{k},z)|\psi(\textbf{k},0)\rangle$, where
 \begin{equation}
     U(\textbf{k},z) = \mathcal{T} e^{-i\int\limits_0^z H(\textbf{k},z')dz{'}}
 \end{equation}
is the evolution operator.  The evolution over each roundtrip period of the microrings is given by the Floquet operator, $U_F(\textbf{k}) = U(\textbf{k},L)$, whose eigenstates are the Floquet modes $|\Phi_n(\textbf{k},0)\rangle$ with eigenvalues $e^{-i\epsilon_n(\textbf{k})L}$. 
In the absence of detuning ($\Delta\phi = 0$), the quasienergy bands $\epsilon_n(\textbf{k})$ of the Floquet modes form composite transmission band manifolds, each containing $4N^2$ degenerate bulk modes and separated by bandgaps.  The effect of the phase detune $\Delta\phi$ is to break the degeneracy and lift one Floquet mode into the bandgap, forming an isolated single band (Fig.~\ref{fig:2}(a)). Moreover, this energy-shifted band becomes increasingly flattened as the phase detune is increased, implying that the field distribution becomes more strongly localized spatially.  Figures~\ref{fig:2}(b)-(d) show the field distributions of the isolated Floquet mode when each of microrings $A$, $C$ and $D$, respectively, is detuned during step $j = 1$. When microring $A$ is detuned, the field is localized in two coupled bulk-mode loops sharing the common segment $j = 1$. By contrast, detuning microring $C$ results in the field strongly localized in only a single bulk-mode loop traced out by the hopping sequence.  A similar mode pattern is also observed when segment $j = 2$ of microring $B$ is detuned.  When the weakly-coupled microring $D$ is detuned, light does not follow the driving sequence but instead remains trapped in the same site resonator. This point-defect mode is also observed in a topologically trivial lattice (see Supplemental Material). Thus by selectively applying phase detunes to specific steps in the driving sequence, distinct mode patterns can be excited to probe the topological behavior of the Floquet lattice.

\begin{figure}[t]
    \centering
    \includegraphics[width=1\linewidth]{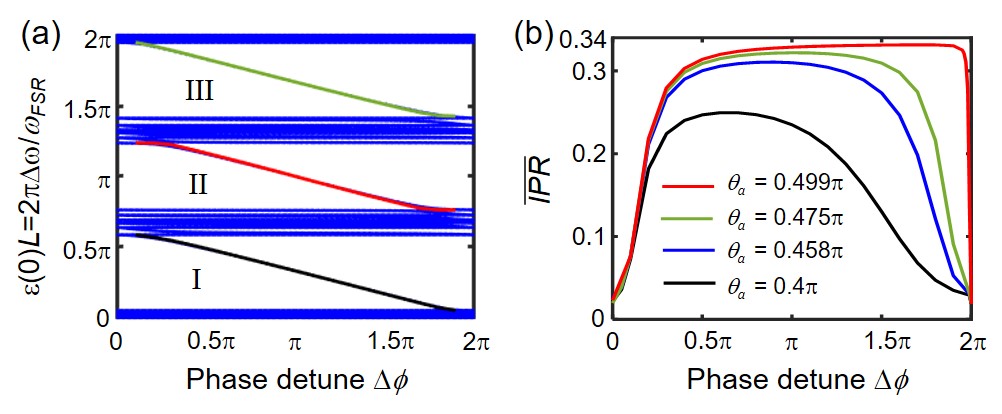}
    \caption{Effects of the phase detune on the resonant frequency and spatial localization of FMR. (a) Dependence of the cyclic phase change  $\epsilon_s(\textbf{0})L$ and resonant frequency shift of the FMR in each bandgap of an AFI lattice ($\theta_a = 0.458\pi$, $\theta_b=0.025\pi$) on the phase detune $\Delta\phi$.  The blue lines are the quasienergies of the transmission bands, which remain largely unchanged with phase detuning. (b) Variation of the average inverse participation ratio of FMR (in bandgap III) with phase detune $\Delta\phi$ for Floquet lattices with coupling angle $\theta_a$ varied from $0.4\pi$ to $0.499\pi$ and $\theta_b$ fixed at $0.025\pi$.
    }
    \label{fig:3}
\end{figure}

The strong field localization in a bulk-mode loop is effectively a resonance effect caused by the energy-shifted Floquet mode constructively interfering with itself after completing each roundtrip around the loop.  
Starting out each cycle at $z = 0$, the shifted Floquet mode $|\Phi_s(\textbf{k},0)\rangle$ evolves as $|\Psi_s(\textbf{k},z)\rangle = U(\textbf{k},z)|\Phi_s(\textbf{k},0)\rangle$. According to Floquet theorem, the state $|\Psi_s(\textbf{k},z)\rangle$ can also be expressed as \cite{Nakagawa}
\begin{equation}
    |\Psi_s(\textbf{k},z)\rangle = e^{-i\epsilon_s(\textbf{k})z}|\Phi_s(\textbf{k},z)\rangle
\end{equation}
where $|\Phi_s(\textbf{k},z)\rangle = e^{i\epsilon_s(\textbf{k})z}U(\textbf{k},z)|\Phi_s(\textbf{k},0)\rangle$ is the periodic $z$-evolved Floquet state satisfying $|\Phi_s(\textbf{k},z + L)\rangle = |\Phi_s(\textbf{k},z)\rangle$. Thus, the state $|\Psi_s(\textbf{k},z)\rangle$ will constructively interfere with itself after every period $L$ if
 \begin{eqnarray}
    |\Psi_s(\textbf{k},z+L)\rangle &=& e^{-i\epsilon_s(\textbf{k})(z+L)}|\Phi_s(\textbf{k},z+L)\rangle \nonumber \\
    &=&e^{-i\epsilon_s(\textbf{k})z}|\Phi_s(\textbf{k},z)\rangle
\end{eqnarray}
from which we obtain the condition for constructive interference as $\epsilon_s(\textbf{k})L = 2m\pi$, $m \in \mathbb{Z}$. Using the quasienergy for a stationary Floquet mode at $\textbf{k} = \textbf{0}$, we can calculate the shift in the resonant frequency of the FMR relative to a microring resonance as $\Delta\omega_s = \epsilon_s(\textbf{0})L \Delta\omega_{FSR}/2\pi$, where $\Delta\omega_{FSR}$ is the free spectral range (FSR) of the microrings. Figure~\ref{fig:3}(a) plots the dependence of the cyclic phase change $\epsilon_s(\textbf{0})L$ on the phase detune $\Delta\phi$, showing that the resonant frequency of an FMR can be continuously tuned across a topological bandgap. In the modern theory of electric polarization in crystals, the Berry phase of a Bloch mode (divided by $2\pi$) represents the averaged displacement from a lattice site of the corresponding Wannier state in real space \cite{Marzari2012}. By the same analogy, the cyclic phase $\epsilon_s(\textbf{k})L$, which includes both the dynamical and geometric phases of the system, can be interpreted as the average position of the Floquet mode in the frequency domain \cite{Shem2019,Nakagawa}.  Here we show that by tuning the cyclic phase of a Floquet-Bloch mode, we can vary its frequency position to yield a resonance localized in both spatial and frequency domains in an otherwise homogeneous topological lattice.

We note that this resonance effect is cavity-less since it does not require physical boundaries between the lattice and another medium but instead relies on an adiabatic change in the Hamiltonian via a phase detune.  Since no interface scattering takes place, FMRs can in principle have very high Q factors.  Importantly, since the phase detune $\Delta\phi$ represents a local adiabatic change to the Hamiltonian $H_{FB}$, the energy-shifted band still retains the topological properties of the unperturbed lattice. This is evident from the fact that the FMR mode (Fig.~\ref{fig:2}(c)) retains the same spatial distribution of a bulk mode in a homogeneous lattice as we increase the phase detune. Also, the bandgaps above and below the FMR still support edge modes, implying that the topological behavior of the lattice is not altered by the adiabatic phase detuning. In particular, simulations show that the FMR remains robust to random variations in the lattice (see Supplemental Material).

We can quantify the degree of spatial localization of an FMR by computing its inverse participation ratio (IPR) \cite{bell1970atomic}. For a $z$-evolved Floquet mode with normalization $\langle{\Phi_s(\textbf{0},z)|\Phi_s(\textbf{0},z)\rangle}=1$, we can define the average IPR over one evolution period as
\begin{equation}
    \overline{IPR}= \frac{1}{L} \sum_{k = 1}^{4N^2}\int\limits_0^L |\Phi_s^{(k)}(\textbf{0},z)|^4 dz 
\end{equation}
 where $\Phi_s^{(k)}$ is the field in site resonator $k$ in the lattice. Figure~\ref{fig:3}(b) shows the average IPR of an FMR (in bandgap III) as a function of the phase detune for different coupling angles of the lattice. It is seen that the mode becomes more strongly localized as it is pushed deeper into the bandgap. Thus, in general, we can expect to achieve the strongest intensity enhancement for FMRs located near the center of the bandgap.  The degree of localization is also higher for lattices with larger contrast between the coupling angles $\theta_a$ and $\theta_b$.  We note that the maximum $\overline{IPR}$ achievable for FMR is 1/3 because at any given position $z$ in an evolution cycle, the field is localized in three separate microrings in the bulk-mode loop.

\section{Experimental Demonstration of FMR}
 
\begin{figure*}[t]
    \centering
    \includegraphics[width=1\linewidth]{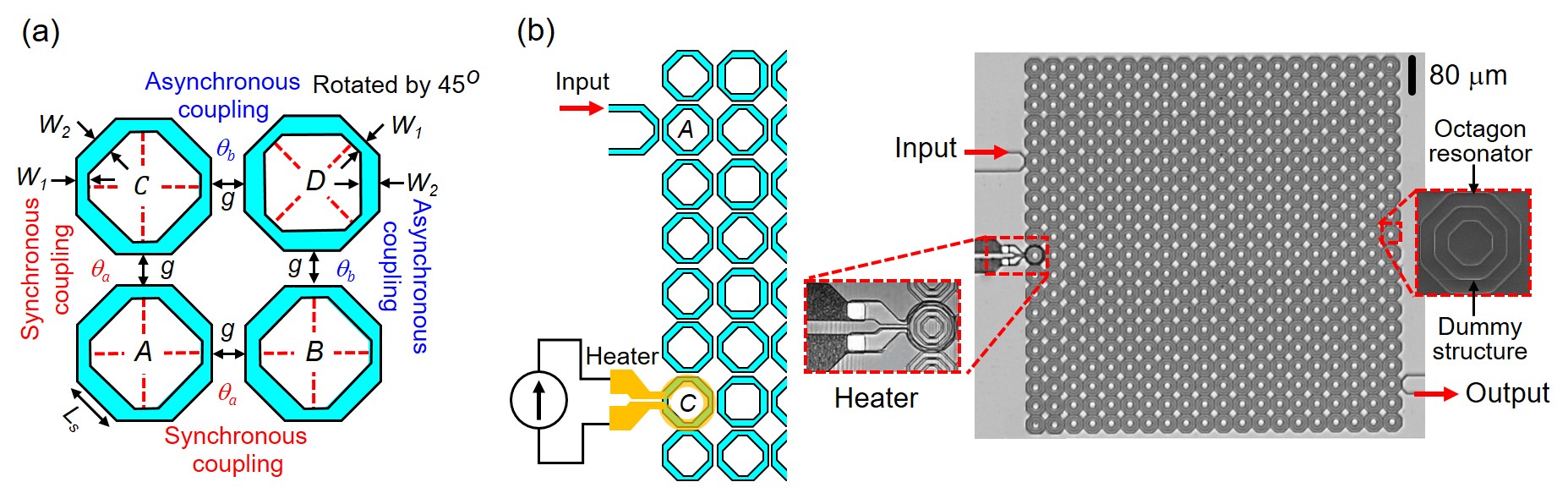}
    \caption{Design and implementation of the Floquet octagon lattice in SOI for demonstrating FMR. (a) Schematic of a unit cell of the Floquet octagon lattice, with octagon $D$ rotated by $45^o$ with respect to octagons $A$, $B$ and $C$ to realize synchronous and asynchronous coupling angles $\theta_a$ and $\theta_b$. (b)  Microscope image of the fabricated octagon lattice in SOI showing the input and output waveguides used to measure the transmission spectrum, and the heater used to tune the phase of an octagon $C$ on the left boundary to excite FMR.}     \label{fig:4}
\end{figure*}

We demonstrated FMR in a Floquet TPI lattice consisting of a square array of coupled octagon resonators on an SOI substrate.  Each unit cell consisted of 4 identical octagons evanescently coupled to their neighbors via identical coupling gaps $g$ (Fig.~\ref{fig:4}(a)).  Octagon resonators were used to realize dissimilar coupling angles $\theta_a$ and $\theta_b$ in each unit cell by exploiting the difference between synchronous and asynchronous couplings. This is achieved by designing the octagons to have sides with equal lengths $L_s$ but alternating widths $W_1$ and $W_2$, with octagon $D$ rotated by $45^o$ with respect to the other 3 octagons. Octagon $A$ can thus be strongly coupled to its neighbors via synchronous coupling between waveguides of the same width $W_1$, while octagon $D$ is weakly coupled to its neighbors via asynchronous coupling between waveguides of dissimilar widths $W_1$ and $W_2$.  We designed the coupling angles to be $\theta_a=0.458\pi$ and $\theta_b =0.025\pi$, so that the lattice exhibits AFI behavior for TE polarized light in its three bandgaps over one FSR at the telecommunication wavelengths \cite{afzal2020realization} (see Supplemental Material  for details of the lattice design). The fabricated lattice consisted of $10\times10$ unit cells (Fig.~\ref{fig:4}(b)). An input waveguide was coupled to resonator $A$ of a unit cell on the left boundary of the lattice to excite AFI edge modes and an output waveguide was coupled to resonator $B$ on the right boundary to measure the transmission spectrum.

\begin{figure*}[htp]
    \centering
    \includegraphics[width=1\linewidth]{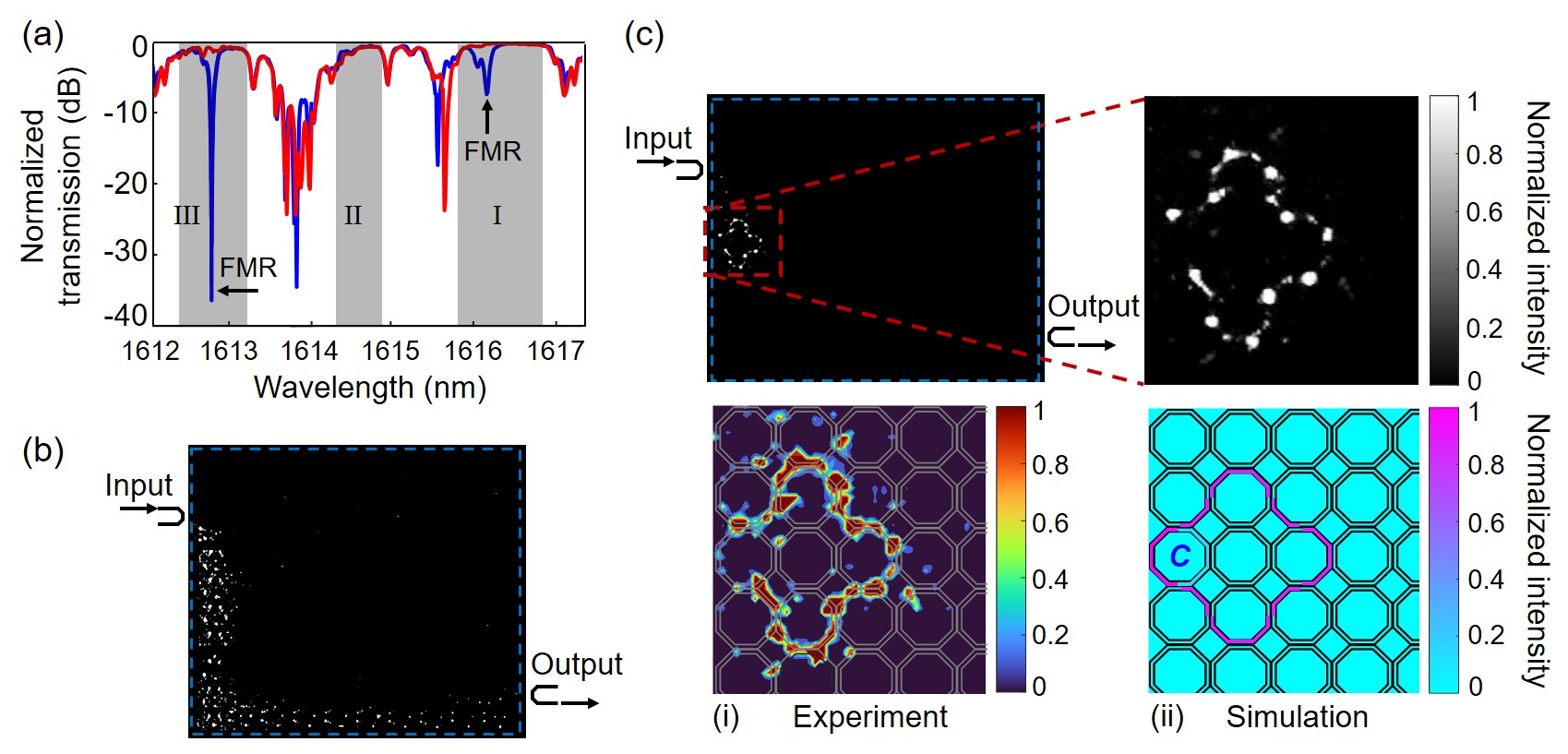}
    \caption{Experimental observation of FMR. (a) Measured transmission spectra of the Floquet octagon lattice over one FSR when there was no phase detune (red trace) and when a phase detune of $\Delta\phi = 1.45\pi$ was applied to microring $C$ on the left boundary (blue trace). (b) NIR camera image of scattered light intensity at 1612.833 nm wavelength in bandgap III with no phase detune, showing an AFI edge mode propagating along the left and bottom edges of the lattice.  (c) NIR image at 1612.833 nm wavelength with phase detune of $\Delta\phi = 1.45\pi$, showing FMR localized in a bulk-mode loop. The edge mode is not visible due to its much weaker intensity compared to the FMR. Inset (i) shows a map of scattered light intensity reconstructed from raw camera data superimposed on the octagon lattice; inset (ii) shows the simulated intensity distribution of the FMR for comparison.}
    \label{fig:5}
\end{figure*}

Figure~\ref{fig:5}(a) (red trace) shows the transmission spectrum measured for input TE light over one FSR ($\sim$ 5 nm) of the resonators around 1615 nm wavelength (see Supplemental Material for measurement setup).  Three distinct bands of high transmission (labeled I, II, III) due to AFI edge mode propagation can be seen, which correspond to the topologically nontrivial bulk bandgaps of the Floquet lattice.  Imaging of the scattered light intensity distribution at $1612.833$ nm wavelength (in bandgap III) using an NIR camera (Fig.~\ref{fig:5}(b)) confirms the formation of an edge mode propagating along the lattice boundary from the left input waveguide to the right output waveguide.

\begin{figure*}[t]
    \centering
    \includegraphics[width=1\linewidth]{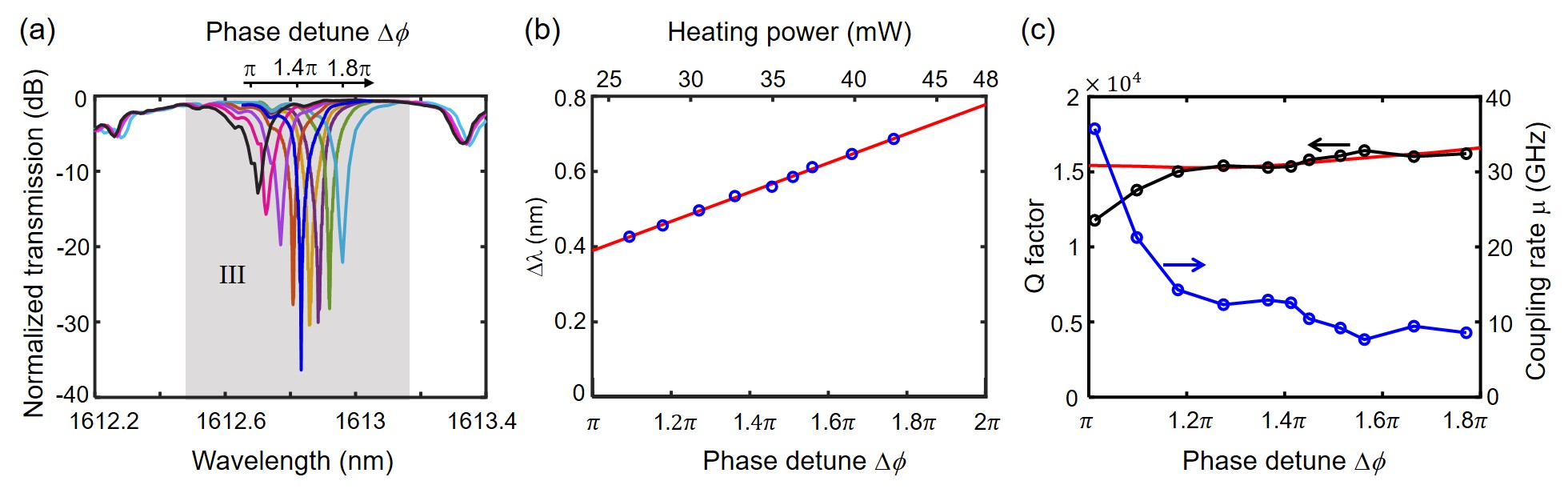}
    \caption{Tuning of FMR across the topological bandgap:  (a) transmission spectra of FMR in bandgap III at various phase detunes. Top horizontal scale indicates the phase detunes $\Delta\phi$ corresponding to the resonance dips. (b)  Dependence of the resonant wavelength shift of the FMR (relative to the microring resonance at zero phase detune) on the phase detune (bottom horizontal axis) and heating power (top horizontal axis). Blue circles are measurement data; red line is the linear best fit. (c) Variations of the extrinsic Q factor and the coupling rate $\mu$ of the FMR with phase detune $\Delta\phi$.  Black circles are measured Q; red line is the simulated Q of FMR in a lattice with $\theta_a = 0.458\pi$, $\theta_b = 0.025\pi$ and roundtrip loss of 0.59 dB in each octagon.}
    \label{fig:6}
\end{figure*}

Since the FMR exists in a bulk bandgap, we can couple light into it using an AFI edge mode in the same bandgap.  To excite an FMR near the left boundary of the lattice,
we thermo-optically tuned the phase of an octagon $C$ on the left boundary (Fig.~\ref{fig:4}(b)) using a titanium-tungsten heater fabricated on top of the resonator (detuning the whole resonator $C$ at the edge also excites only a single bulk-mode loop).
Figure~\ref{fig:5}(a) (blue trace) shows the transmission spectrum when a phase detune of $1.45\pi$ (corresponding to electrical heating power $P= 34.9$ mW) was applied to the octagon (see Supplemental Material  for heater calibration method).  We observe that the spectrum is almost identical to the spectrum without phase detune (red trace), except for the presence of two sharp dips located in bulk bandgaps I and III. These dips indicate the presence of an FMR excited in each bulk bandgap by the edge mode. To obtain visual confirmation of the spatial localization of the FMR, we performed NIR imaging of the scattered light intensity at the resonance wavelength $1612.833$ nm (in bandgap III) (Fig.~\ref{fig:5}(c)). The image clearly shows that light is localized and trapped in a bulk-mode loop, which is not present in Fig.~\ref{fig:5}(b) when no phase detune was applied.  The bulk mode pattern directly captures the hopping sequence of the Floquet lattice as predicted in Fig.~\ref{fig:1}(c). Strikingly, the edge mode does not ``go around" the detuned octagon $C$ as when it encounters a defect, but instead excites the FMR and couples to it.
We also note that transmission dips occurring in the bulk transmission bands of the lattice, which appear with and without phase detuning, are caused by random interference of light propagating deep into the lattice bulk.  Imaging of light intensity patterns at these wavelengths in the transmission band does not show light localized in FMR loops \cite{afzal2020realization}.

Focusing on the FMR in bandgap III, we measured the resonance spectrum for different phase detune values.  The spectra are plotted in Fig.~\ref{fig:6}(a), showing that as the phase detune is increased, the FMR spectrum is pushed deeper into the bandgap.  The resonance linewidth also becomes narrower while the extinction ratio reaches a maximum of almost $-40$ dB near the bandgap center.  In Fig.~\ref{fig:6}(b) we plotted the resonant wavelength shift $\Delta\lambda$ as a function of the phase detune, with the corresponding heater power shown on the top horizontal axis.  The linear relationship between $\Delta\lambda$ and $\Delta\phi$ is in agreement with the theoretically predicted dependence of the FMR quasienergy on the phase detune (Fig.~\ref{fig:3}(a)).

The dependence of the Q factor of the FMR on the phase detune is shown in Fig.~\ref{fig:6}(c) (black circles). We obtained Q values in the range $1.2 \times 10^4 - 1.6 \times 10^4$, with a slight increasing trend as the FMR moves deeper into the bandgap.  For comparison, the intrinsic Q factor of a single resonator obtained from measurement of a stand-alone octagon (see Supplemental Material) was only slightly higher at $2.6 \times 10^4$ (corresponding to roundtrip loss of 0.35 dB).    Using the designed coupling values ($\theta_a = 0.458\pi$, $\theta_b = 0.025\pi$) for the lattice and a slightly higher roundtrip loss of 0.59 dB in each octagon, we simulated FMR spectra for various phase detunes and obtained the corresponding extrinsic Q factors (red line in Fig.~\ref{fig:6}(c)), which show good agreement with the measured values.
Larger discrepancies between simulated and measured Q factors are observed for smaller phase detunes, which can be attributed to the fact that the FMR and edge mode are less localized near the band edge and are thus more susceptible to lattice imperfections.  From the measured extrinsic Q factors $Q_{ex}$, we can calculate the effective coupling ($\mu$) between the FMR and the edge mode as $\mu = \omega_0(1/Q_{ex}-1/Q_0)$, where $\omega_0$ is the resonant frequency and $Q_0 = 1.85\times10^4$ is the intrinsic Q factor. The results are also plotted in Fig.~\ref{fig:6}(c) (blue circles).  The coupling rate $\mu$ depends on the overlapping between the field distributions of the AFI edge mode and the FMR.  This dependence is seen to correlate with the variation in the degree of spatial localization of the FMR as indicated by the plot of $\overline{IPR}$ vs. $\Delta\phi$ in Fig.~\ref{fig:3}(b).  As the FMR is pushed deeper into the bandgap, it becomes more strongly localized spatially so that its coupling to the edge mode is weaker, which results in higher Q factor.  
Using the designed coupling angle values, simulations of the lattice showed that the theoretical extrinsic Q factor (without loss) can exceed $10^6$, suggesting that the experimental Q factor can be improved by reducing the roundtrip loss in the FMR loop, for example, by reducing scattering from the octagon corners and using materials with lower absorption.

\section{Discussion and Conclusion}

The reported FMR represents a new mechanism for trapping light in a TPI lattice by adiabatically tuning the cyclic phase of a Floquet mode to induce quantum self-interference.  The spatial localization pattern of the FMR is shown to be dictated by the driving sequence of the Floquet lattice and is thus a manifestation of its topological nature. Our work thus introduces a simple way for directly exciting and isolating a Floquet bulk mode for probing its topological properties.  The nontrivial behavior of the lattice can be deduced from the bulk-mode hopping pattern, in contrast to the approach used in previous works on TPIs, which typically rely on imaging edge modes at the lattice boundaries to verify its topological nature. We note that it is also possible to perturb the coupling angles in the driving sequence, which may provide additional mechanism for probing the bulk modes and controlling the FMRs.

Compared to other topological resonators, the FMR is cavity-less, tunable, and can be formed anywhere in the lattice bulk. The lack of physical cavity boundaries suggests that very high Q factors can potentially be achieved. The resonance can also be dynamically switched on and off, which could be useful for realizing optical switches and modulators.  In addition, our preliminary experimental results have shown that multiple adjacent FMRs could be excited to form coupled cavity systems, which could open up new applications such as high-order coupled-cavity filters, optical delay lines, and light transport in a bandgap through the lattice bulk by hopping between adjacent localized bulk modes.

\section*{Acknowledgments}
This work was supported by the Natural Sciences and Engineering Research Council of Canada.

%%%%%%%%%% If using BibTeX:
%\bibliography{reference}

%%%%%%%%%% If preparing manually:

\end{document}

% This section should spread over both columns

\preprint{APS/123-QED}

\title{Supplemental Material: Floquet Mode Resonance : Trapping light in the bulk mode of a Floquet topological insulator by quantum self-interference}% Force line breaks with \\

\author{Shirin Afzal}
%\email{safzal1@ualberta.ca}

\author{Vien Van}%
\affiliation{Department of Electrical and Computer Engineering, University of Alberta, Edmonton, AB, T6G 2V4,
Canada}

\date{\today}% It is always \today, today,

\maketitle

%\tableofcontents

\section{ Comparison between  FMR and defect state in a topologically trivial Floquet lattice}

\begin{figure}[h]
    \centering
    \includegraphics[width=0.65\linewidth]{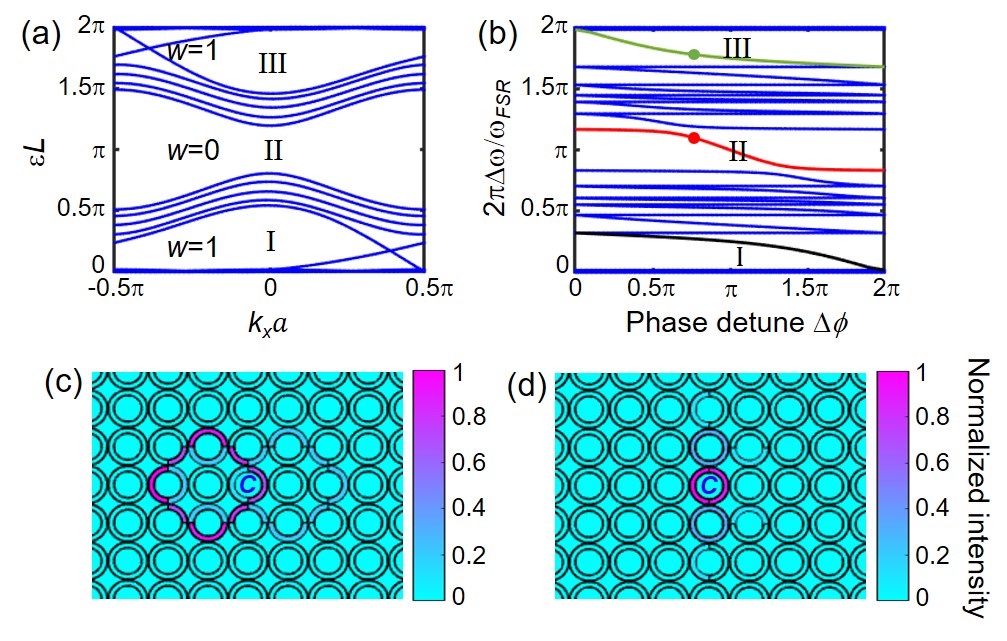}
    \caption{Comparison between FMR and point-defect state in a Floquet microring lattice. (a) Projected quasienergy band diagram of a microring lattice with coupling angles $\theta_a=0.3\pi$ and $\theta_b=0.01\pi$, with 5 unit cells in the $y$ direction and infinite extent in $x$ ($a$ is the spacing between adjacent microrings). Bandgaps I and III are nontrivial with winding number $w = 1$; bandgap II is trivial with $w = 0$. (b) Quasienergies of the Floquet states  as functions of phase detune  $\Delta\phi$ applied to step $j = 1$ of a microring $C$ in a lattice with 5 $\times$ 5 unit cells. The shifted states are indicated by the green, red and black curves. (c) and (d) Intensity distributions of energy-shifted bulk modes in bandgaps III and II, respectively, when phase detune $\Delta\phi=0.75\pi$ is applied to the lattice. These states are indicated by the green and red dots in (b).}
    \label{fig:S1}
\end{figure}

Here we provide a contrasting example of the different behaviors of an FMR in a topologically nontrivial bandgap and a conventional defect state in a trivial bandgap, both existing in the same Floquet lattice. We consider a microring lattice with coupling angles $\theta_a=0.3\pi$ and $\theta_b=0.01\pi$, which behaves as a Floquet Chern insulator in bandgaps I and III and as a normal insulator in bandgap II (see Ref. \cite{afzal2018topological} for the topological characterization of Floquet microring lattices).  These behaviors can be verified by the projected band diagram of a semi-infinite lattice (with 5 unit cells in the $y$ direction and infinite extent in the $x$ direction) in Fig.~\ref{fig:S1}(a), which shows edge states existing in bandgaps I and III but not in bandgap II.  We apply a phase detune $\Delta\phi$ in step $j=1$ of a microring $C$ in the lattice, which shifts the quasienergy of a bulk mode from each transmission band manifold into the bandgap below, as shown in Fig.~\ref{fig:S1}(b).  Although the trivial bandgap II hosts an energy-shifted bulk state, the spatial field distribution of the mode is markedly different from those in the nontrivial bandgaps I and III.  For instance, for the same phase detune of $\Delta \phi=0.75\pi$, the intensity distributions of the shifted states in bandgaps III and II are shown in Figs.~\ref{fig:S1}(c) and ~\ref{fig:S1}(d), respectively.  The shifted bulk mode in the nontrivial bandgap is not localized in the detuned resonator but also hops to neighbor sites, forming a loop defined by the hopping sequence which gives rise to the nontrivial topological behavior of the Floquet lattice.  On the other hand, the shifted bulk mode in the trivial bandgap is localized in the same detuned site resonator, which is similar to a point-defect mode in a static lattice. Thus it is possible to determine the topological behaviour of a lattice by probing the spatial distribution of the energy-shifted bulk mode in its bandgap.  This technique complements the conventional approach of exciting edge modes along the lattice boundaries to verify its nontrivial topological behavior.

\section{Robustness of FMR}

\begin{figure} [h]
    \centering
    \includegraphics[width=1\linewidth]{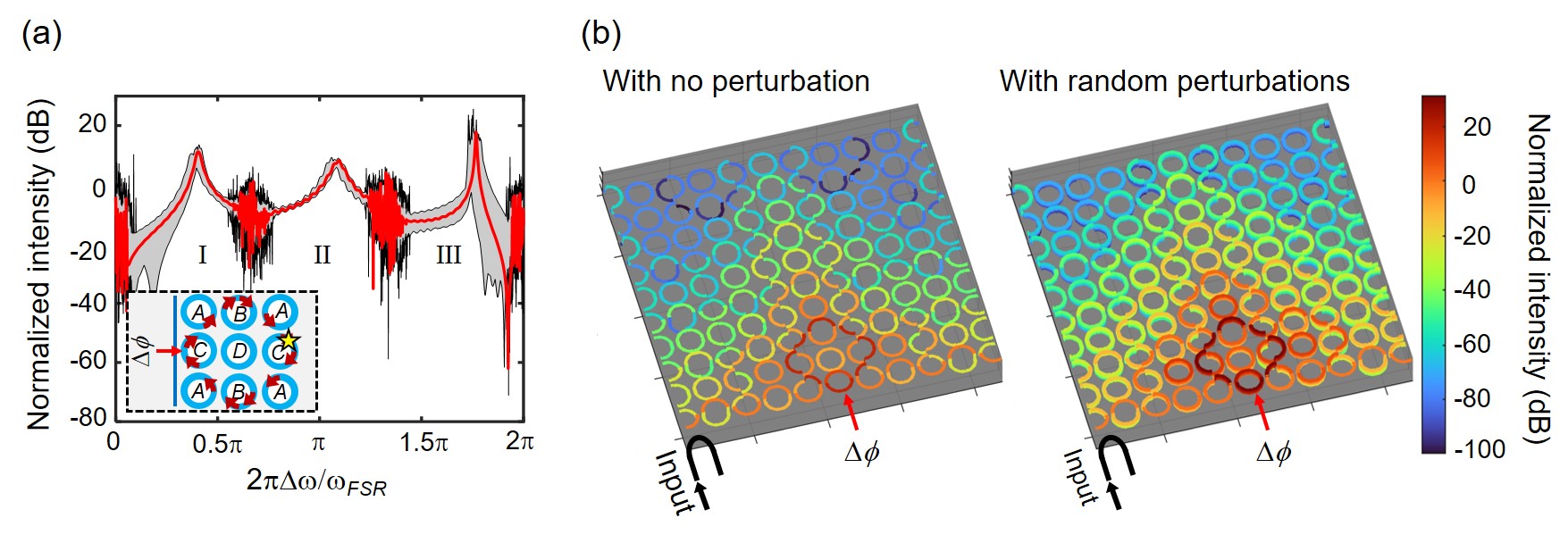}
    \caption{
   Effects of random variations in the Floquet microring lattice on FMR: (a) Simulated spectral responses of light intensity inside the FMR (at the location indicated by yellow star in the inset figure), when a phase detune $\Delta\phi=1.45\pi$ is applied to microring $C$ on the left boundary. Red trace is the ideal lattice with no perturbation; the grey area shows the variations in the intensity due to $\pm 10\%$ random perturbations in the coupling angles and microring roundtrip phases obtained from 100 simulations. (b) Intensity distributions of the FMR in bandgap III without and with the random perturbations (obtained from 20 simulations). The FMR loop appears in dark red color, which is also clearly visible for the case with random perturbations, suggesting that the spatial localization of the FMR is also robust to variations.}
    \label{fig:S2}
\end{figure}

We investigated the robustness of FMR in the presence of random variations in the Floquet microring lattice shown in Fig. 4(b) of the main text, with $10\times 10$ unit cells and coupling angles $\theta_a=0.458\pi$, $\theta_b=0.025\pi$.  We formed an FMR near the left boundary by applying a phase detune of $\Delta\phi=1.45\pi$ to a microring $C$ on the left boundary. We coupled light into the FMR using AFI edge mode, which is excited through the input waveguide.  Figure~\ref{fig:S2}(a) shows the simulated spectral response of light intensity inside the FMR loop (at the location indicated by the yellow star in the inset diagram in Fig.~\ref{fig:S2}(a)). The red trace is the ideal case with no random variations in the lattice, showing three resonant peaks appearing in the 3 bandgaps over one FSR of the microrings.  The grey area indicates the variations in the intensity due to uniformly-distributed random deviations of up to $\pm 10\%$ in the coupling angles and roundtrip phases of the microrings in the lattice. It is seen that the FMR peaks still appear in the 3 bandgaps at approximately the same quasienergies, implying that the frequency position of an FMR is robust to random variations.
Figure~\ref{fig:S2}(b) compares the spatial distributions of light intensity of the FMR in bandgap III without and with the random variations.  It is seen that while random variations cause light to be spread out more to the resonators surrounding the FMR loop, most of the light is still strongly localized in a bulk-mode loop.  Thus the spatial localization of an FMR is also robust to random variations.

\section{Design and measurement of Floquet octagon lattice}

The Floquet octagon lattice was realized on an SOI substrate with 220 nm-thick Si layer lying on a 2 $\mu$m-thick SiO$_2$ buffer layer with a 2.2 $\mu$m-thick SiO$_2$ overcladding layer.  The octagon resonators had sides of length $L_s = 16.08$ $\mu$m and alternating widths $W_1 = 400$ nm and $W_2 = 600$ nm.  The corners were rounded using arcs of radius $R$ = 5 $\mu$m to reduce scattering.  The coupling gaps between adjacent octagons were fixed at $g = 225$ nm. From numerical simulations using the Finite-Difference Time-Domain solver in Lumerical software \cite{lumerical2018solutions}, we obtained
$\theta_a = 0.458\pi$ and $\theta_b = 0.025\pi$ for the synchronous and asynchronous coupling angles, respectively, around 1615 nm wavelength. To excite AFI edge mode and measure the transmission spectrum, input and output waveguides with 400 nm width were coupled to their respective octagon resonators on the left and right boundaries of the lattice via the same coupling gap $g = 225$ nm.  The $10 \times 10$ lattice was fabricated using the Applied Nanotools SOI process \cite{ANT}.

Transmission spectra of the lattice were measured using a tunable laser source (Santec TSL-510 $1510-1630$ nm), with the polarization adjusted to TE before being coupled to the input waveguide through a lensed fiber.  The transmitted light was also collected with a lensed fiber for measurement with a photodetector and power meter.  Imaging of the scattered light intensity patterns from the chip was performed using a $20 \times$ objective lens and an InGaAs NIR camera. The image sensor of the camera had $640 \times 512$ pixels with a $15$ $\mu$m pitch. 

We excited an FMR and tuned its resonant frequency by varying the phase of an octagon $C$ on the left boundary of the lattice.  A TiW heater covering the octagon perimeter (Fig. 4(b) in the main text) was fabricated on top of the resonator and current was applied to the heater to tune its phase. Figure 6(b) in the main text shows that the measured resonant wavelength shift $\Delta\lambda$ of the FMR varies approximately linearly with the applied heater power $P$.  Using fact that the roundtrip phase detune $\Delta\phi$ of the octagon resonator also varies linearly with the heater power, we can correlate the measured $\Delta\lambda$ vs. $P$ plot with the simulated $\Delta\lambda$ vs. $\Delta\phi$ plot across the bandgap. This allows us to calibrate the heater efficiency and deduce the linear correspondence between the phase detune and the heater power. The relationship between $\Delta\phi$ and $P$ is explicitly shown on the top and bottom horizontal axes of Fig. 6(b) in the main text.

\section{Loss measurement of a stand-alone octagon resonator}

\begin{figure}[h]
    \centering
    \includegraphics[width=1\linewidth]{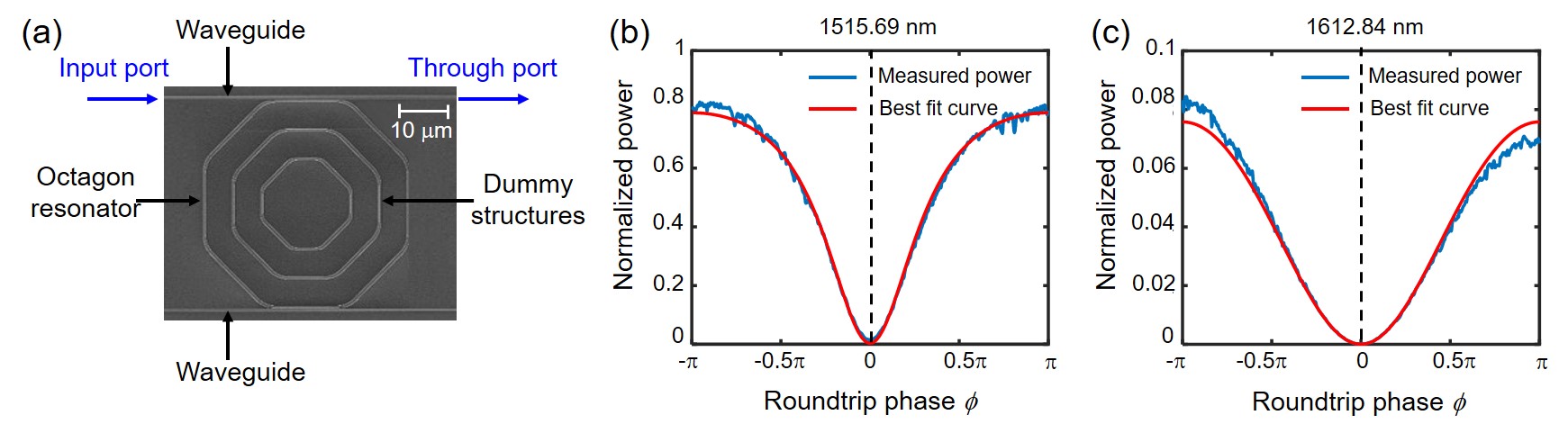}
    \caption{Characterization of a stand-alone octagon resonator. (a) Image of a stand-alone octagon resonator coupled to two waveguides along the top and bottom sides. (b) and (c) Measured power transmission spectra at the through port and best curve fits of resonances around 1515 nm and 1615 nm wavelength, respectively.   }
    \label{fig:S3}
\end{figure}

To determine the loss in the octagon resonators in the Floquet lattice, we fabricated a stand-alone octagon resonator with the same dimensions as the octagons in the lattice.  The octagon was evanescently coupled to two waveguides of 400 nm width at the top and bottom sides, as shown in Fig.~\ref{fig:S3}(a).  The coupling gaps between the resonator and the waveguides were $g=225$ nm, which is the same as those between adjacent octagon resonators in the Floquet lattice.  We measured the transmission spectrum of the resonator at the through port by scanning input TE-polarized light over the 1510 nm - 1630 nm wavelength range.  Two sample resonance spectra around 1515 nm and 1615 nm wavelengths are shown in Figs.~\ref{fig:S3}(b) and ~\ref{fig:S3}(c), respectively.  To determine the coupling coefficients and loss of the resonator, we fit each resonance spectrum using the equation for the power transmission at the through port \cite{van2016optical}:
\begin{equation}
    T_t=\frac{T_{min}+F\sin^2(\phi/2)}{1+F\sin^2(\phi/2)}
    \label{Eq:modeling_stand_alone}
\end{equation}
where $T_{min} = \frac{\tau^2(1 - a_{rt})^2}{(1 - \tau^2a_{rt})^2}$ and $F=\frac{4\tau^2a_{rt}}{(1-\tau^2a_{rt})^2}$. In these expressions, $\tau$ is the transmission coefficient of the coupling sections, $\phi$ is the roundtrip phase and $a_{rt}$ is the roundtrip amplitude attenuation constant in the resonator.  Sample best fit curves are shown for the two spectra in Figs.~\ref{fig:S3}(b) and ~\ref{fig:S3}(c).  From the curve fits, we obtained $a_{rt} =0.96$, which corresponds to an intrinsic Q factor of $2.6\times 10^4$ for the stand-alone resonator. The coupling angle $\theta=cos^{-1}(\tau)$ could also be obtained from the curve fit.  For the two sample spectra, we obtained $\theta=0.286\pi$ around 1515 nm and $\theta=0.453\pi$ around 1615 nm.  The latter value is in good agreement with the designed synchronous coupling angle $\theta_a$ of the Floquet lattice around the same wavelength.

\newpage

% Bibliography
%\bibliography{Ref.bib}